\documentclass{ws-mpla}

\usepackage{graphicx}
\usepackage{dcolumn}
\usepackage{bm}

\begin{document}

\markboth{K.-C. Lai \& P. Chen}
{Influence of Plasma Collective Effects on Cosmological Evolution}


\title{Influence of Plasma Collective Effects on Cosmological Evolution}

\author{Kwang-Chang Lai\footnote{\email{kclai@mail.nctu.edu.tw}}}
\address{Institute of Physics, National Chiao Tung University,
Hsinchu, 300 Taiwan.\\
Leung Center for Cosmology and Particle Astrophysics, National Taiwan University,\\
Taipei, 106 Taiwan.}

\author{Pisin Chen\footnote{\email{chen@slac.stanford.edu}}}
\address{Kavli Institute for Particle Astrophysics and Cosmology, Stanford Linear Accelerator Center, Stanford University,
Stanford, California 94309, USA\\
Department of Physics and Graduate Institute of Astrophysics, and Leung Center for Cosmology and Particle Astrophysics, National Taiwan University,
Taipei, 106 Taiwan.}%

\date{\today}

\maketitle

\begin{abstract}

It is well-known that the universe was in a plasma state both before decoupling and after reionization. However, the conventional wisdom has been that the plasma effects are largely Debye-shielded and can thus be safely ignored when considering large scale evolutions. Recently we showed that large scale structure formation in the universe may actually be suppressed by the plasma collective effect. Indeed, observational data indicate that the conventional theoretical formula tends to overestimate the matter power spectrum at scales $k>1h{\rm Mpc}^{-1}$. In this paper, we further develop our theory through a more thorough and general derivation of the Maxwell-Einstein-Boltzmann equation. In addition to baryon density perturbation post reionization, we apply this general formulation to investigate the possible plasma effect on CMB anisotropy. As expected, while the plasma effect does render an observable effect to the former, its impact on the latter is totally negligible.


\end{abstract}

\ccode{PACS Nos.:98.65.Dx, 95.30.Qd, 98.80.Bp}

\section{Introduction}

Recently we showed that large scale structure formation in the universe would be suppressed by the plasma collective effect\cite{CnL}. Indeed, observational data seem to indicate that the conventional theoretical formula tends to overestimate\cite{Teg} the matter power spectrum at scales $k>1h{\rm Mpc}^{-1}$. It is implied that the underlying theory may have to be modified. Moreover, our work\cite{CnL} is performed in the regime of Newtonian limit and we expect to generalize the plasma collective effect  in the general relativistic regime. 

On the other hand, the cosmological perturbation theory(CPT) is not complete in the sense that the electromagnetic interaction is absent. Einstein-Boltzmann equations accounting for evolution of cosmological perturbations have been explored for decades\cite{Lifshitz,PnY,BnS}. Concerning anisotropies, the Sachs-Wolfe effect\cite{SW} and Silk damping\cite{Silk} were predicted just after the discovery of CMB\cite{3K} and the approach of tightly coupled limit\cite{DZS} was perfected\cite{Seljak,HnS1995} in 1990's to account for the CMB spectrum. Neither these studies include EM fields even though  tightly coupled limit approach combines photons and baryons as a cosmic plasma, nor have investigations been performed on the likeliness of plasma effects in CMB study. In parallel, the efforts on matter inhomogeneities including study on the growth and transfer functions\cite{Blu,Pee,HnS1996} comprise the evolution under only gravity. Later modification\cite{EnH} taking baryons into account neither addressed EM interactions, though it derived more accurate fitting formula to replace the pure cold dark matter BBKS fitting\cite{BBKS}. Though the CPT as well as its predictions and description of CMB and MPS do not take the EM interaction into account, we expect plasma collective effects for the universe was in plasma state before decoupling and after reionization.

In the paper, we generalize the Einstein-Boltzmann(EB) equations in the CPT to our Maxwell-Einstein-Boltzmann(MEB) equations, which describe the perturbation evolution under both the gravity and EM interaction. With the inclusion of plasma effects we derive not only CMB ion-acoustic oscillations but also baryon ion-acoustic oscillations.  The paper is organized as follow. In Sec. II, we justify that the plasma conditions apply in the universe during pre-decoupling and post-reionization era. Based on this justification, we generalize the EB equation to  the MEB equation in Sec.III. The CMB and baryon ion-acoustic wave equations are then derived, respectively, in Sec.IV and V. Finally, we discuss the significance of plasma effects on cosmological perturbation.

\section{Plasma State of the Universe}

By saying "The universe is in plasma state", we mean that relevant entities in the universe satisfy plasma conditions and hence the universe can be modelled as a plasma. As is said that a plasma is a quasineutral gas of charged and neutral particles which exhibits collective behavior\cite{Chen},  two criteria--quasineutrality and collectiveness-- must be fulfilled for the universe. As a perfectly uniform plasma is charge neutral and has no net
electrostatic potential due to the perfect Debye shielding, the inevitable thermal fluctuations of
its electron density at finite temperature would, however, induce
a non-vanishing residual electrostatic potential of the ion that
defies the Debye shielding. Such a residual electric force exerts
an additional pressure on ions (baryons) which will be incorperated into the CPT to display the ion-acoustic behavior in the MEB equation in the following sections. The quasineutrality now means that the scale of interests (i.e. the size of the universe) is much larger than than the Debye length:
\begin{equation}
\lambda\sim 1/k \sim 10^{28}(1+z)^{-1}\hspace{.5mm}{\rm cm}\gg\lambda_D\equiv\sqrt{\gamma_e k_B T_e/4\pi e^2 n_e}.
\end{equation}
This makes the system is neutral enough so that one can take $n_i=n_e$ and not so neutral that all interesting EM interactions vanish. In addition, collective behaviors require enough particles inside the Debye sphere:
\begin{equation}
N_D= n\frac{4}{3}\pi\lambda_D^3 \gg 1.
\end{equation}

Before decoupling ($z\gtrsim 1000$), $T_e=T_\gamma=T_{\gamma,0}(1+z)$ and $n_e\sim 10^{-10}n_\gamma=10^{-10}n_{\gamma,0}(1+z)^3$ with the present photon temperature and number density $T_{\gamma,0}=2.73K$ and $n_{\gamma,0}\simeq 400{\rm cm}^{-3}$ and redshift $z>1000$. These information renders
\begin{eqnarray}
\lambda_D &=& 1.5\times 10^4 (1+z)^{-1} \hspace{.1cm} {\rm cm},\\
N_D          &=& 5.7\times 10^5 \hspace{.1cm} {\rm cm}^{-3}
\end{eqnarray}

After reionization, $T_e=T_b=T^*_b(1+z)^2/(1+z^*)^2$. As indicated observationally, the redshift of reionization $z^*$ ranges as $5\lesssim z^*\lesssim 20$ \cite{Becker,Fan,Kogut,Spergel} and the reionization temperature $T_b^*$ is at least higher than $1000K$. Similarly,
\begin{eqnarray}
\lambda_D &=& 1.6\times 10^4 T_b^* (1+z)^{-1/2}(1+z^*)^{-1}  \hspace{.1cm} {\rm cm} ,\\
N_D          &=& 6.8\times 10^5 T_b^*  (1+z)^{3/2}(1+z^*)^3  \hspace{.1cm} {\rm cm}^{-3}
\end{eqnarray} 

The plasma conditions are apparently satified in both pre-decoupling and post-reionization era and, therefore, our plasma treatment is applicable.

\section{The Maxwell-Einstein-Boltzmann Equation}

Before decoupling, the EB equations can be derived systematically\cite{Dodel} to account for the couplings between cosmological components. The equation displaying the acoustic feature on CMB spectrum is then rendered from these coupled equation. Following this systematical way, our MEB equation is deduced by introducing the EM force appropriately and combining it with gravity for the Boltzmann equation of baryons.

The most general form of Boltzmann equation is
\begin{equation}
\frac{df}{dt}=C[f], \label{Boltzmann}
\end{equation}
where $f$ is the distribution function and $C[f]$ characterizes
the collision terms. In CPT, the distribution function $f$ can be
expressed as $f(t,E(p),\hat{p}^i,x^i)$ with energy $E$ and
momentum $p$, moving in direction $\hat{p}$ at $(x,t)$. For
radiations (relativistic species), $E=p$ and $E^2=p^2+m^2$ for
matter with mass $m$. Since the collective behavior emerges in the EB equation for baryons and photon EB equations remain unchanged, we present only the derivation of modification for baryons. Following the convention as in
Ref. \refcite{Dodel}, the LHS of the Boltzmann equation is written
as
\begin{equation}
\frac{df}{dt}=\frac{\partial f}{\partial t}+\frac{\partial
f}{\partial x^i}\frac{dx^i}{dt}+ \frac{\partial f}{\partial
E}\frac{dE}{dt}+\frac{\partial f}{\partial
\hat{p}^i}\frac{d\hat{p}^i}{dt}\hspace{0.5mm}.\label{bol}
\end{equation}
In equilibrium, the distribution function depends only on the length of the momentum, i.e. the energy $E(p)$ and not on its direction $\hat{p}^i$, so its dependence on $\hat{p}^i$ should emerge from perturbations and be of the order. Therefore $\partial f/\partial \hat{p}^i$ and $\partial\hat{p}^i/\partial t$ are both first oorder terms and the last term in Eq.(\ref{bol}) is negligible in the linear theory.

Choosing the conformal Newtonian Gauge, $ds^2=-(1+2\Psi)dt^2+a^2(1+2\Phi)\delta_{ij}dx^i dx^j$, where $a$ is the scale factor, $\Psi$ the Newtonian gravity and $\Phi$ the space curvature, the constraint $g_{\mu\nu}P^\mu P^\nu=-m_b^2$ renders the four-momentum for baryons
\begin{equation}
P^\mu=\left[E(1-\Psi),p\hat{p}^i\frac{1-\Phi}{a} \right]  \label{4-momen}
\end{equation}
where $P^\mu\equiv dx^\mu/d\xi$, with $\xi$ parametrizing the particle's path.

This definition renders directly 
\begin{equation}
\frac{dx^i}{dt}=\frac{dx^i}{dx^0}=\frac{P^i}{P^0}=\frac{p\hat{p}^i}{aE}(1-\Psi+\Phi).\label{10}
\end{equation}

$dE/dt$ (or $dp/dt$) is deduced from the geodesic equation
\begin{equation}
\frac{d^2 x^\mu}{d\xi^2}=-\Gamma^\mu_{\alpha\beta}\frac{dx^\alpha}{d\xi}\frac{dx^\beta}{d\xi}\hspace{5mm}
{\rm or, equivalently,}\hspace{5mm}\frac{dP^\mu}{d\xi}=-\Gamma^\mu_{\alpha\beta}P^\alpha P^\beta.
\end{equation}
This geodesic can be regarded as a variation of Newton's second law describing the motion under the 
gravity in the form of metric perturbation. Since baryons feel not only gravity but also EM interaction, the geodesic equation should be modified to address the EM field as well. 
The equation of motion under Lorentz' force reads\cite{Rohrlich}
\begin{equation}
a^\mu\equiv\frac{d^2 x^\mu}{d\zeta^2}=\frac{q}{m_b}F^\mu_\alpha\frac{dx^\alpha}{d\zeta}=\frac{q}{m_b}F^\mu_\alpha P^\alpha\frac{d\zeta}{d\xi},
\end{equation}
where $q$ is the charge of the baryon(proton) and $F_{\mu\nu}=\partial_\mu A_\nu-\partial_\nu A_\mu$ is the EM tensor with four-potential $A^\mu$.\\
Here $\zeta$, the proper time, is related to $\xi$ via the constraint of $P^\mu$ as
\begin{equation}
-m_b^2=P^\mu P_\mu=\frac{dx^\mu}{d\xi}\frac{dx_\mu}{d\xi}=\frac{ds^2}{d\xi^2}=-\frac{d\zeta^2}{d\xi^2}.
\end{equation} 
Therefore $d\zeta=m_bd\xi$.\\
By superposition principle, the geodesic equation is modified to address both gravity and EM force 
\begin{equation}
\frac{dP^\mu}{d\xi}=-\Gamma^\mu_{\alpha\beta}P^\alpha P^\beta
                                        +qF^\mu_\alpha P^\alpha. \label{geo0}
\end{equation}

As the Christoffel symbol $\Gamma^\mu_{\alpha\beta}$ is expressed in terms of $g_{\mu\nu}$
\begin{equation}
\Gamma^\mu_{\alpha\beta}=\frac{1}{2}g^{\mu\nu}(\partial_\alpha g_{\nu\beta}+\partial_\beta g_{\alpha\nu}
                                        -\partial_\nu g_{\alpha\beta}),
\end{equation}
the EM tensor $F_{\mu\nu}$, in terms of $A^\mu$, is derived from the electromagnetic lagrangian 
\begin{equation}
\mathcal{L}_{EM}=-\frac{1}{8\pi}F^{\mu\nu}F_{\mu\nu}=\frac{1}{4\pi}(\mathcal{E}^2-\mathcal{B}^2),\label{LEM}
\end{equation}
where $\mathcal{E}$ and $\mathcal{B}$ are electric and magnetic fields induced from the charge density fluctuation, 
and related with electric and magnetic potentials $\phi$ and $\mathcal{A}$ by
$\mathcal{E}=-\nabla\phi$ and $\mathcal{B}=\nabla\times\mathcal{A}$, with the gradient operator $\nabla_i\equiv\frac{\partial}{a\partial x^i}$.

Evaluating Eq.(\ref{LEM}) renders the four-potential
\begin{equation}
A^\mu=\left[\frac{\phi}{a}(1-\Psi+\Phi), \frac{\mathcal{A}\hat{\mathcal{A}}^i}{a}\right].
\end{equation}
As $\mu=0$ designated and $d\xi$ replaced by $dt$, expanding Eq.(\ref{geo0}) renders, with Eq.(\ref{10}), 
\begin{equation}
\frac{df}{dt}=\frac{\partial f}{\partial t}+\frac{\partial f}{\partial x^i}\frac{p\hat{p}^i}{aE}
-\frac{\partial f}{\partial E}\left(H^2\frac{p^2}{E}+\frac{\partial\Phi}{\partial t}\frac{p^2}{E}
+\frac{p\hat{p}^i}{a}\frac{\partial\Psi}{\partial x^i}+q\frac{p\hat{p^i}}{aE}\frac{\partial\phi}{\partial x^i}\right), \label{lhs}
\end{equation}
the LHS of Eq.(\ref{Boltzmann}).

The last term inside the bracket describes the new effect from perturbed electric fields ensured by the total neutrality of the universe while all others are conventional terms. $\phi$, like $\Psi$, is of the first order itself since only its gradient counts.  The absence of vector potentials results straightly from the fact that they change the momentum but the energy of charged particles.  Because of the $\hat{p}^i$ dependence, the extra em-term makes no contribution in the zeroth moment equation, like the curvature term. It contributes in the first moment equation and renders the MEB equation for baryons
\begin{equation}
\dot{v}_b+aHv_b+ik(\Psi +\frac{e\phi}{m_b})=\frac{\dot{\tau}}{R}(3i\Theta_1+v_b),\label{24}
\end{equation}
where "$\cdot$" denotes derivatives with respect to the conformal
time and $R\equiv\frac{3\rho_b}{4\rho_\gamma}$ is the
baryon-photon energy density ratio, $\Theta_1$ the first moment of
the temperature perturbation $\Theta$, and $\tau$ the optical
depth.

Since the perturbed electric field $\phi$ arises from the density perturbation,
\begin{equation}
\frac{e\phi}{m_b}=\frac{1}{m_b}\frac{4\pi
e^2n_0}{k^2+\lambda_D^{-2}}\delta=\frac{k_BT}{m_b}\delta_b=\frac{k_BT}{m_b}3\Theta_0,\label{25}
\end{equation}
where the last equality is enforced by the adiabatic condition.
The MEB equation Eq.(\ref{24}) is then written as
\begin{equation}
v_b = -3i\Theta_1+\frac{R}{\dot{\tau}}\left[\dot{v}_b+aHv_b+ik\left(\Psi+\frac{k_BT}{m}3\Theta_0\right)\right].\label{26}
\end{equation}

\section{CMB Anisotropy Before Decoupling}

The photon distribution function, with the temperature anisotropy $\Theta\equiv\delta T/T$, is
\begin{equation}
f(t,p,\hat{p}^i,x)=\left[exp\left\lbrace\frac{p}{T(t)[1+\Theta(t,\hat{p}^i,x)]}\right\rbrace -1\right]^{-1}.
\end{equation}

Since photons are not disturbed by EM interactions, the photon MEB equation is the same as its EB version
\begin{equation}
\dot{\Theta}+ik\mu\Theta+\dot{\Phi}+ik\mu\Psi=-\dot{\tau}[\Theta_0-\Theta+\mu v_b], \label{28}
\end{equation}
where the moments are defined as, with Legendre polynomials $P_l$,
\begin{equation}
\Theta_l\equiv\frac{1}{(-i)^l}\int^1_{-1} \frac{d\mu}{2}P_l(\mu)\Theta(\mu).
\end{equation}
 In the tightly coupled limit, the only nonnegligible moments are the monopole and dipole. Eq.(\ref{28}) renders the coupled equations of the zeroth and first moments,
\begin{eqnarray}
\dot{\Theta}_0+k\Theta_1=-\dot{\Phi}; \label{30}\\
\dot{\Theta}_1-\frac{k\Theta_0}{3}=\frac{k\Psi}{3}+\dot{\tau}\left[\Theta_1-\frac{iv_b}{3}\right]. \label{31}
\end{eqnarray} 
Eqs.(\ref{26}, \ref{30} and \ref{31}) allow to extract our MEB equation for the temperature anisotropy $\Theta_0$ :
\begin{eqnarray}
\ddot{\Theta}_0&+&aH\frac{R}{1+R}\dot{\Theta}_0
          +k^2\left[\frac{1}{3(1+R)}+\frac{R}{1+R}\frac{k_BT}{m}\right]\Theta_0  \nonumber \\
&=&-\frac{k^2}{3}\Psi-aH\frac{\dot{R}}{1+R}\dot{\Phi}-\ddot{\Phi}
\end{eqnarray}

From this MEB equation, the sound speed of the CMB fluid is changed as
\begin{equation}
c_s^2=\frac{1}{3(1+R)}\rightarrow\frac{1}{3(1+R)}+\frac{R}{1+R}\frac{k_BT}{m_b},
\end{equation}
where the former term represents the conventional theory and the last reveals the plasma collective contribution. The contribution to the sound speed from the EM induced pressure relative to the photon pressure is easily estimated
\begin{equation}
\sqrt{R\frac{k_BT}{m_b}}\sim10^{-5}
\end{equation}
Accordingly, the sound horizon is changed to the same order. The plasma contribution should therefore occur at scales much larger than the Silk damping and is observationally negligible.

\section{Baryon Perturbation After Reionization}

After reionization, the universe is transferred to another plasma state and EM interactions influence the evolution of the baryon density perturbation again. In this epoch, we study the evolution of perturbations, using Newtonian gravity, as in Ref. \refcite{LnL}. Since photons were decoupled long ago, Boltzmann equation in the integral form, the fluid equation is appropriate to describe the density pertubation. With EM forces included, our evolution equation for baryon perturbation is derived from plasma equations in the expanding universe
\begin{eqnarray}
\frac{d\rho_b}{dt}&=&\frac{\partial \rho_b}{\partial t}+({\bf u_b}\cdot\nabla)\rho_b=-3H\rho_b;\label{continuity}\\
\frac{d{\bf u_b}}{dt}&=&-\nabla\Psi_{\rm gr}-\frac{1}{\rho_b}\nabla P-\frac{e}{m}\nabla\phi_{\rm em}, \label{Euler}
\end{eqnarray}
where $\rho_b$ and ${\bf u_b}$ denote the mass density and velocity of the baryon and $\Phi_{\rm gr}$, $P$ and $\phi_{\rm em}$ are the gravity, pressure and em-potentials the baryon suffers. The gravitational potential should account for total mass $\rho_m$ including baryons and the dark matter via Poisson equation
\begin{equation}
\nabla^2\Psi_{\rm{gr}}=4\pi G\rho.
\end{equation}
The universe is charge neutral in total so the electric field arises from the density fluctuation and is perturbative, $\phi_{\rm em}=\phi$. The expansion is defined by the local Hubble parameter as
\begin{equation}
H=\frac{1}{3}(\nabla\cdot {\bf u_b}).
\end{equation}
Perturbing and linearizing these equations render, to first order,
\begin{eqnarray}
(\delta\rho_b)^\cdot&=&-3\rho_b\delta H-3H\delta\rho_b,\label{lin-1}\\
\dot{v}_b+Hv_b &=&-\nabla\Psi-\frac{1}{\rho_b}\nabla\delta P
                     -\frac{e}{m_b}\nabla\phi,\\\label{lin-2}
          \delta H&=&\frac{1}{3}\nabla\cdot v_b,\label{lin-3}\\
      \nabla^2\Psi&=&4\pi G\delta\rho,
\end{eqnarray}
where $"\cdot"$ denotes partial differentiation with respect to time $t$.
Instead of $\delta\rho_b$, the density contrast $\delta_b\equiv\delta\rho_b/\rho_b$ is more useful to work with. The MEB equation is then extracted from these coupled equations
\begin{equation}
\ddot{\delta}_b+2H\dot{\delta}_b-\frac{1}{\rho}\nabla^2\delta P
-\frac{e}{m_b}\nabla^2\phi=4\pi G\rho\delta_b. \label{54}
\end{equation}
Under Fourier transformation, our MEB equation for baryons in reionization era is  
\begin{equation}
\ddot{\delta}_{{\bf k},b}+2H\dot{\delta}_{{\bf k},b}+\Big(\frac{k}{a}\Big)^2(1+\frac{5}{3})\frac{k_BT}{m_b}\delta_{{\bf k},b}=4\pi G\rho\delta_{{\bf k},b},
\end{equation}
where thermodynamics leads the $\delta P$-term to
\begin{equation}
\frac{\delta P_k}{\rho_b}=\frac{\delta\rho_{{\bf k},b}}{\rho_b}\frac{\delta
P_{\bf k}}{\delta\rho_{{\bf k},b}}=\frac{5k_BT}{3m_b}\delta_{{\bf k},b}.
\end{equation}

We therefore deliver the origin of our MEB equation, according to which our investigation of plasma suppression was performed in Ref. \refcite{CnL}.

\section{Conclusion}

The acoustic behavior is an impressive feature in the CMB
spectrum and matter power spectrum. Our study shows manifestly the EM fields also contribute to this
acoustic oscillation collectively. Though the whole system is neutral in total, the nonzero temperature induced charge density fluctuations form a collective electric field which turns to be an extra portion of pressure exerting on the densithy perturbations in both cases. However, the
contribution on the CMB is too small to be observed. Both the order of magnitude
estimation and the ion fluid equation analysis tell that the EM
contribution is of the same order as of the pressure from charged
particles. As the adiabatic condition indicates baryon and photon density contrasts being parallel, the baryon to photon number ratio $n_b/n_\gamma\sim10^{-10}$ states that the EM effect on the sound speed is $\sim10^{-5}$ smaller than
ordinary CMB fluctuation.  Our generalization of the Boltzmann equation proves this number by
incorporating the Lorentz force into the ordinary geodesic
equation for baryons.

Though the plasma collective effect is negligibly small in CMB spectrum, it should has sizable imprints in the matter power spectrum as discussed in our work\cite{CnL}.

\section*{Acknowledgements}

KCL 
is supported by the National Science Council (NSC
096-2811-M-009-024) of Taiwan, R.O.C.
PC is supported in part by US Department of Energy under Contract No.\
DE-AC03-76SF00515 and in part by the National Science Council (NSC 96-2811-M002-001) of Taiwan R.O.C.

\end{document}